\documentclass[twocolumn]{article}
\usepackage{graphicx,epsfig,color}
\usepackage[normal]{subfigure}
\usepackage{amsmath,amsthm,amssymb}
\usepackage{floatflt}  
\usepackage{setspace}
\usepackage{rotating,algorithm,algorithmic}

\usepackage{xcolor,colortbl}
\definecolor{red}{rgb}{0.82,0.1,0.26}
\definecolor{green}{rgb}{0.4,1.0,0.0}
\definecolor{light-gray}{gray}{0.79}
\definecolor{dark-gray}{gray}{0.00}

\title{\huge\bf Group-wise classification approach to improve  Android malicious apps detection accuracy}
\author{Ashu Sharma$^1$, Sanjay K. Sahay$^1$\\
\medskip {\small\it(Sanjay K. Sahay)}~\\
{\normalsize Birla Institute of Technology and Science, Pilani, Department of Computer Science and Information Systems,$^1$}\\
{\normalsize Goa Campus, NH-17B, By Pass Road, Zuarinagar-403726, Goa, India}\\
{\normalsize (Email: ssahay@goa.bits-pilani.ac.in)}\\
}

\date
\newpage
\begin{document}
\maketitle

\begin{abstract}
In the fast-growing smart devices, Android is the most popular OS, and due to its attractive features, mobility, ease of use, these devices hold sensitive information such as personal data, browsing history, shopping history, financial details, etc. Therefore, any security gap in these devices means that the information stored or accessing the smart devices are at high risk of being breached by the malware.  These malware are continuously growing and are also used for military espionage, disrupting the industry, power grids, etc. To detect these malware, traditional signature matching techniques are widely used. However, such strategies are not capable to detect the advanced Android malicious apps because malware developer uses several obfuscation techniques. Hence, researchers are continuously addressing  the security issues in the Android based smart devices. Therefore, in this paper using \textit{Drebin} benchmark malware dataset we experimentally demonstrate how to improve the detection accuracy by analyzing the apps after grouping the collected data based on the permissions and achieved 97.15\% overall average accuracy. Our results outperform the accuracy obtained without grouping data (79.27\%, 2017),  Arp, et al. (94\%, 2014), Annamalai et al. (84.29\%, 2016), Bahman Rashidi et al. (82\%, 2017)) and  Ali Feizollah, et al. (95.5\%, 2017). 
The analysis also shows that among the groups, {\it Microphone} group detection accuracy is least while {\it Calendar} group apps are detected with the highest accuracy, and for the best performance, one shall take 80-100 features.
\vspace*{0.1cm}
~
\end{abstract}

\section{Introduction}
The attractive features and mobility of smart devices have drastically changed the today's environment. Many functionalities of these devices are similar to the traditional information technology system, which can also access enterprises applications and data, enabling employees to do their work remotely. Hence the security risks are not only limited to Bring Your Own Smart Device (BYOSD) scenarios but also for the devices which are adopted on an ad hoc basis. Therefore, any security gap in these devices means that the information stored  or accessing smart devices are at high risk of being breached. 
The recent attack shows that the security features in these devices are not as par to completely stop the adversary \cite{sym2017}. Hence smart devices are becoming an attractive target for the online criminal, and they are investing more and more for the sophisticated attacks viz. ransomware or to steal the valuable personal data from the user device. 
\par In the smart devices, Android  is the most popular operating systems and are connected through the internet accessing billions of online websites (an estimate shows that 5 out of 6 mobile phones are working on Android OS \cite{sym042016}).
Its popularity is basically due to its open source, exponential increase in the Android supported apps, third-party distribution, free rich SDK and the very much suited Java language. In this growing Android apps market, it is very hard to know which apps are malicious. 
As per Statista \cite{statista}, there are approximately two million apps at the \textit{Play Store} of Google and also many third-party apps available for the Android users. Hence potential of the malicious apps or malware entering these systems is now at never seen before levels, not only to the normal users but  also for military espionage, disrupting the industry, power grids (e.g., Duqu, StuxNet), etc. \cite{sharma2014evolution}.  In this, Quick Heal Threat Research Labs in the 3rd quarter of 2015 reported that they had received $\sim4.2 \times 10^5$ malware per day for the Android and Windows platforms \cite{qr2015}.  
\par To detect the malware, traditional approaches are based on the signature matching, which is efficient from a time perspective but not relevant for the detection of advanced malicious apps and continuously growing zero-day malware attack \cite{mcafee122016}. 
Also, to evade the signature-based techniques, malware developer uses several obfuscation techniques.
However, to detect the Android malicious apps, time to time, a number  of static and dynamic methods have been proposed  \cite{arp2014drebin}, \cite{rashidi2017Android}, \cite{narayanan2016adaptive}, \cite{feizollah2017androdialysis}.
But, it appears that the proposed methods are not good enough to effectively detect the advanced malware \cite{sharma2014evolution} in the fast-growing internet and Android based smart devices usage into our daily life. Hence researchers are continuously addressing the security issues in the Android based smart devices. 
Therefore, in this paper, for the effective detection of Android malicious apps with high accuracy, we classified the apps after grouping  the collected data based on permissions. The remaining paper is organized as follows. In next Section, we discuss the related work. Section 3 describes how the collected Android apps are grouped, Section 4 explains the feature selection approach, while Section 5 describes our approach for the effective detection of Android malicious apps and the obtained experimental results. Finally, Section 6 contains the conclusion. 

\section{Related work}

In both the two main methods (static and dynamic) used for the classification of malicious apps, selected classifiers are trained with a known dataset to differentiate the benign and malicious apps.
In this, Arp. et al. achieved 94\% detection accuracy by generating a joint vector space using AndroidManifest.xml file and the disassembled code \cite{arp2014drebin}. Seo, et al. also used the same static features viz. permissions, dangerous APIs, and keywords associated with malicious behaviors to detect potential malicious apps  \cite{seo2014detecting}. 
Based on a set of characteristics derived from binary and metadata Gonzalez, et al. proposed a method \textit{DroidKin}, which can detect the similarity among the apps under various levels of obfuscation \cite{gonzalez2014droidkin}. 
Quentin et al., uses op-code sequences to detect the malicious apps. However, their approaches are not suitable to detect the malware which are completely different \cite{jerome2014using}. 

\par In 2015, Smita Naval, et al. proposed an approach by quantifying the information-rich call sequences to detect the malicious binaries and claimed that the model is less vulnerable to call-injection attacks \cite{naval2015employing}.
In 2016, Jae-wook jang, et al. proposed \textit{Andro-Dumpsys}, a hybrid malware detection approach based on the similarity between the malware creator-centric and malware-centric information. Their experimental analysis shows that {\it Andro-Dumpsys} can classify the malware families with good True Positive ($TP$) and True Negative ($TN$), and are also capable of identifying zero-day threats \cite{jang2016andro}. 
Luca Caviglione, et. al. obtained 95.42\% accuracy using neural networks and decision trees  \cite{naval2015employing}.
Sanjeev Das, et al. proposed {\it GuardOl} (a hardware-enhanced architecture), a combined approach using processor and field programmable gate array for online malware detection. Their approach detects 46\% of malware for the first 30\% of execution, while 97\% on complete execution \cite{das2016semantics}. Saracino, et al.,  proposed a host-based malware detection system called MADAM which simultaneously analyzes and correlates the features at four levels  to detect the malware \cite{saracino2016madam}.
Gerardo Canfora, et al. analyzed two methods to detect Android malware, first was based on Hidden Markov Model, while the 2nd one exploits structural entropy and found that the structural entropy can identify the malware family more correctly \cite{canfora2016hmm}. 
Annamalai et al. proposed {\it DroidOl} for the effective online detection of malware using passive-aggressive classifier and achieved an accuracy of 81.29\% \cite{narayanan2016adaptive}.

\par Recently in 2017, Ali Feizollah, et al. evaluated the effectiveness of Android Intents (explicit and implicit) as a distinguishing feature for identifying malicious applications. They conducted experiments using a dataset containing 7406 applications comprising 1846 clean and 5560 infected applications. They achieved the detection rate of 91\% using Android Intent and 83\% using Android permission. With the combination of both the features, they have achieved 95.5\% detection rate \cite{feizollah2017androdialysis}. 
Nikola et. al. estimated F-measure ({\it does not take account of correctly classified benign apps}) of 95.1\% and 89\% by classifying the apps based on source code and permission respectively \cite{milosevic2017machine}.
Bahman Rashidi et al. experimented with the \textit{Drebin} benchmark malware dataset and shown that their model can accurately assess the risk levels of malicious applications and provide adaptive risk assessment based on user input and can find malware with the maximum accuracy of 82\% \cite{rashidi2017Android}.

\section{Grouping of Android Apps}

In Android, apps run as a separate  process with unique user/group ID and operate in an application sandbox so that apps execution can be kept in isolation from other apps and the system. Hence, to access the user data/resources from the system, apps need additional capabilities that are not provided by the basic sandbox. To access data/resources which are outside of the sandbox, the apps have to explicitly request the needed permission. Depending on the sensitivity of data/area, requested permission may be granted automatically by the system or ask the user to approve or reject the request. In Android, these permissions  can be found in Manifest.permission file e.g. 
to use the call service in an Android app, it should specify:\\
\\
{\small
\noindent  $<manifest xmlns:$ \\$Android = ``http://schemas.Android.com/apk/res/Android"$\\
$package="com.Android.app.callApp" > $\\
$<uses-permission Android:$ \\$name = ``android.permission.CALL\_PHONE" /> \\
    ...\\
</manifest>$\\ }

\par In total there are 235 permissions out of which 163 are hardware accessible and remaining are for user information access \cite{olmstead2016apps}. In terms of security, all these permissions can be put into two categories i.e. normal and dangerous permissions \cite{Androperr}. Therefore it will be important to study the classification of Android malicious apps after grouping them into dangerous and normal/other permissions (Table \ref{table:permissions}). Hence in this paper to improve the overall average detection accuracy of Android malicious apps we use \textit{Drebin} \cite{arp2014drebin} 5531 benchmark malware dataset and 4235 benign apps available at Google play store. 
Our analysis shows that the \textit{Drebin} dataset does not contain any apps which need body sensors permission. 
Therefore we ignored the Sensors group in our experimental analysis and made total nine groups (eight groups of dangerous permissions and one group of normal/other permissions) for the detection of Android apps.

\begin{table}[H]
\centering
\small
\begin{tabular}{|l| l|}
\hline  
 Group & \hspace{1.30cm} Permissions\tabularnewline 
\hline 
Calendar & Read calendar and write calendar.\tabularnewline
\hline 

Camera & Use camera.\tabularnewline
\hline 

Contacts & Read contacts, write contacts and \tabularnewline

 & get contacts.\tabularnewline
\hline
Location & Access fine location and \tabularnewline
 & Access coarse location.\tabularnewline
\hline 

Microphone & Record audio.\tabularnewline
\hline 
Phone & Read phone state, call phone, \tabularnewline
& read call logs, add voicemail,\tabularnewline
 &  use sip and process outgoing calls. \tabularnewline
\hline 
Sensors & Use body sensors\tabularnewline
\hline 
SMS & Send SMS, receive SMS, read SMS\tabularnewline

 & receive WAP push and receive MMS.\tabularnewline
\hline 
Storage & Read external storage and \tabularnewline
 & write external storage.\tabularnewline
\hline 
\end{tabular}

\caption{Dangerous permissions groups of the Android apps }
\label{table:permissions}
\end{table}

\section{Feature Selection} \label{drataprocessing}
For the detection of Android malicious apps, feature selection plays a vital role, not only to represent the target concept but also to speed-up the learning and testing process. In this, often datasets are represented by many features. However, few of them may suffice to improve the concept quality, and also limiting the features will speed-up the classification. The Android apps can be represented as a vector of 256 opcodes \cite{opcodeList}, and some of these opcodes can be used as features for the effective and efficient detection of Android  malicious apps.  Therefore, to find the prominent features which can represent the target concept, opcodes from the collected Android apps are extracted as follows
\begin{itemize}
  \item The $.apk$ files (Android apps) has been decompiled by using freely available \textit{apktool}.
  \item From the decompiled data, we kept only the $.smali$ files and discarded other data, and then
  \item Opcodes are extracted from the $.smali$ files.
\end{itemize}

  \begin{figure}[H]
    \centering
        \includegraphics[width=3.0in]{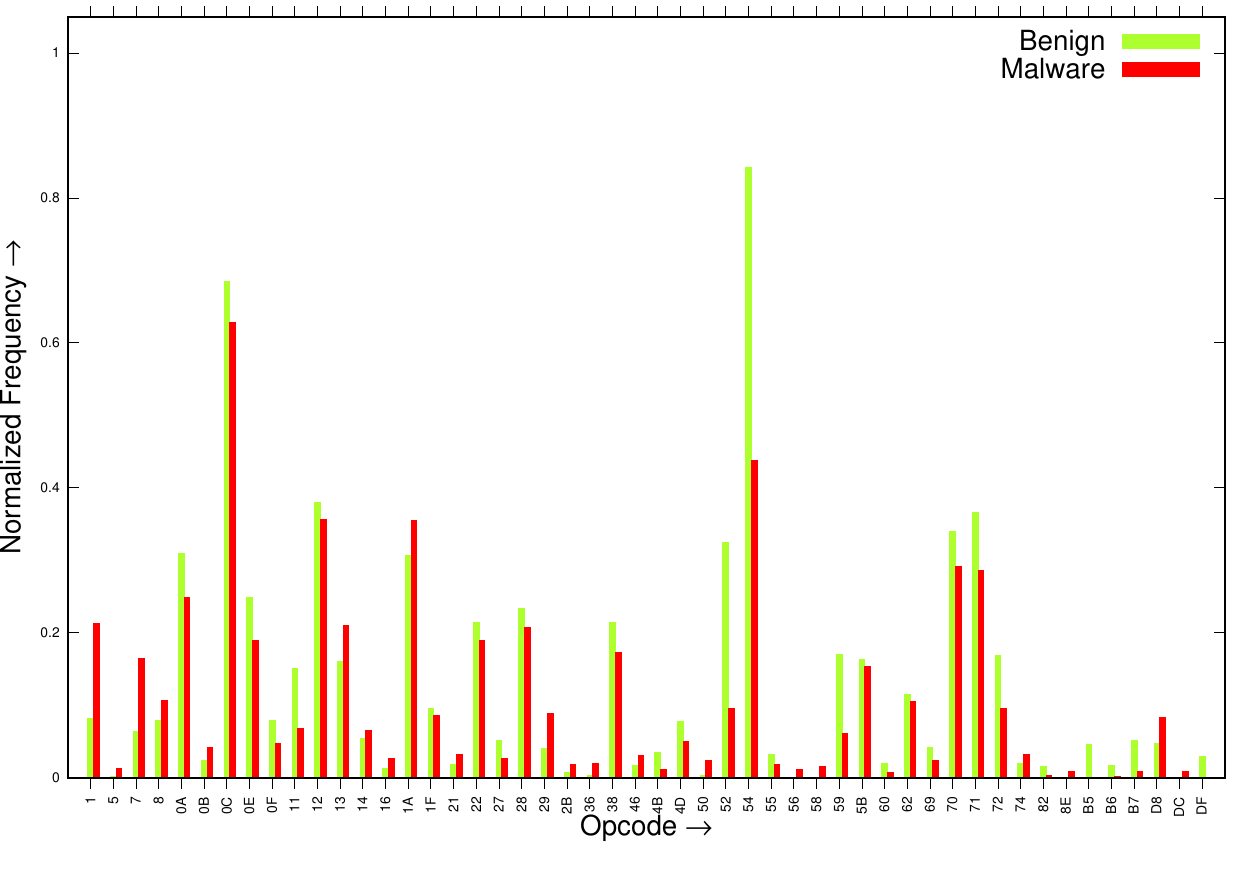} 
\caption{\small \sl Top 50 opcodes occurrence difference between benign and malicious apps in the CALENDAR group.} 
\label{fig:profileg1}
  \end{figure}
   \begin{figure}[H]
    \centering
        \includegraphics[width=3.0in]{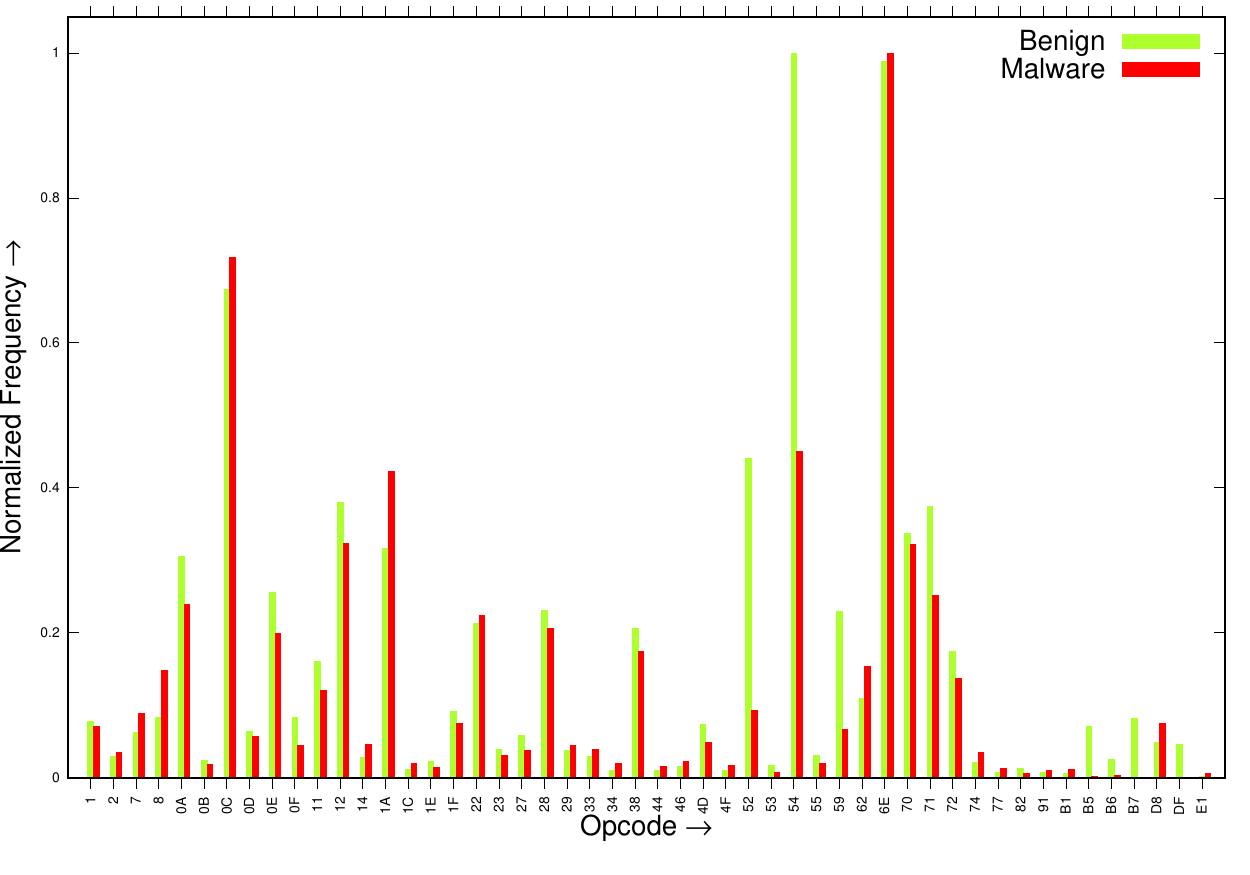} 
\caption{\small \sl Top 50 opcodes occurrence difference between benign and malicious apps in the Camera group.} 
\label{fig:profileg2}
  \end{figure}
    
   \begin{figure}[H]
    \centering
        \includegraphics[width=3.0in]{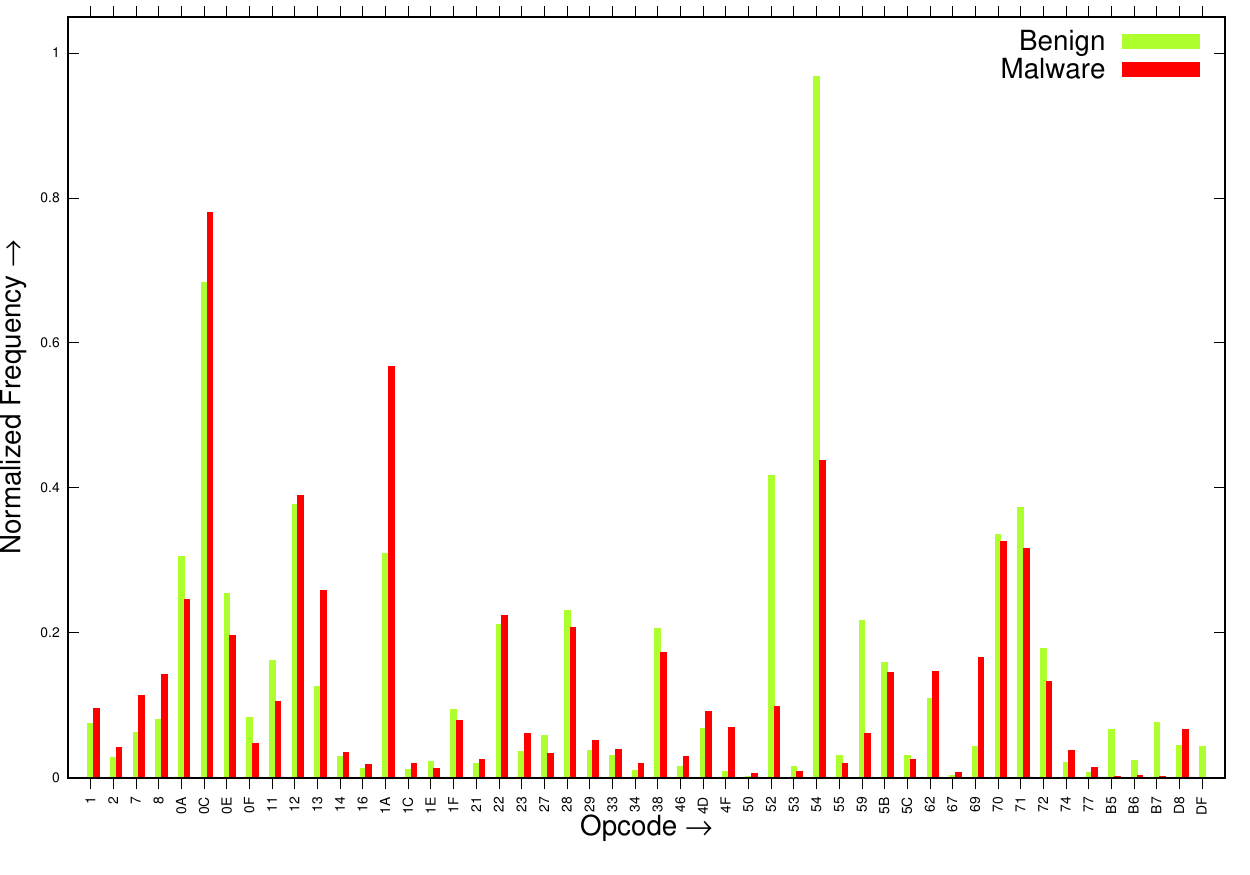} 
\caption{\small \sl Top 50 opcodes occurrence difference between benign and malicious apps in the COTACTS group.} 
\label{fig:profileg3}
  \end{figure}
   \begin{figure}[H]
    \centering
        \includegraphics[width=3.0in]{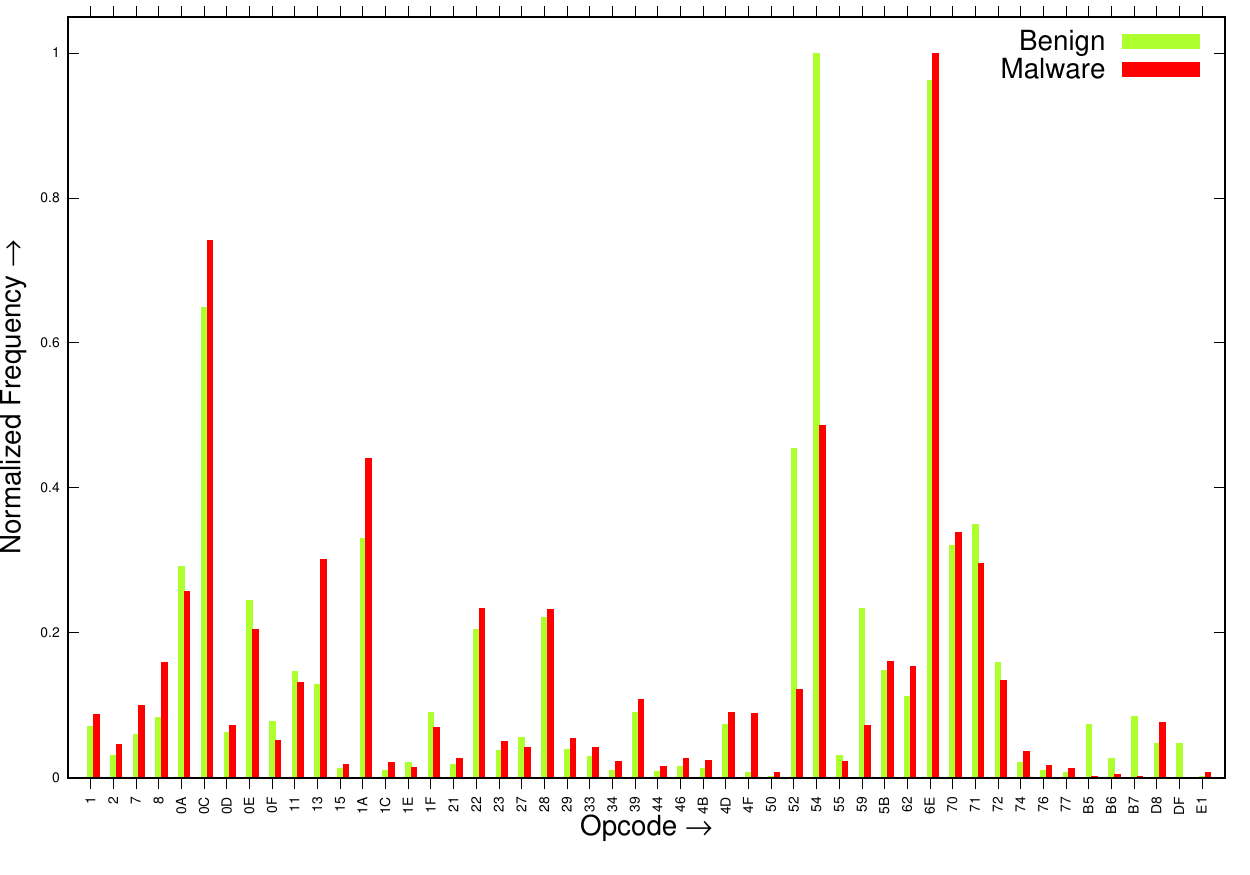} 
\caption{\small \sl Top 50 opcodes occurrence difference between benign and malicious apps in the Location group.} 
\label{fig:profileg4}
  \end{figure}

   \begin{figure}[H]
    \centering
        \includegraphics[width=3.0in]{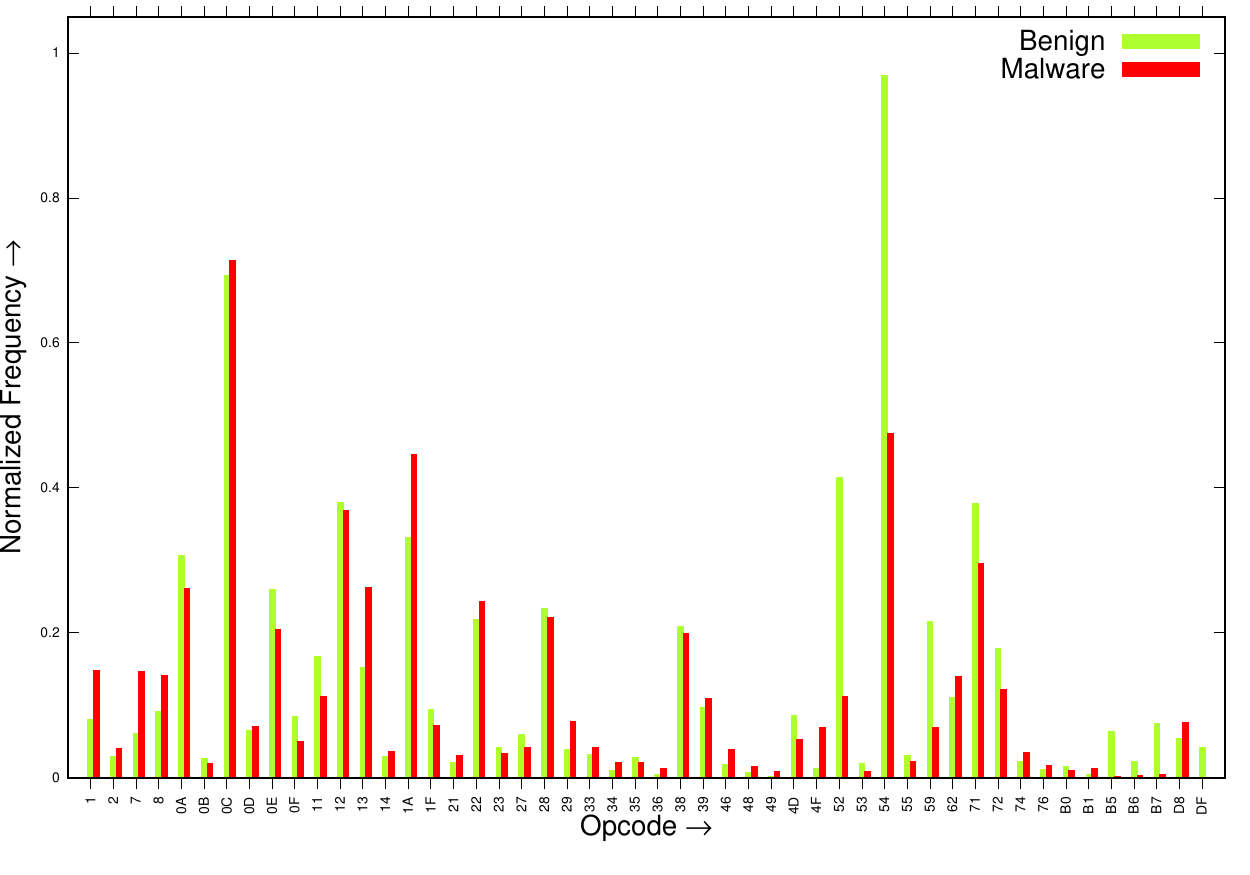} 
\caption{\small \sl Top 50 opcodes occurrence difference between benign and malicious apps in the Microphone group.} 
\label{fig:profileg5}
  \end{figure}

   \begin{figure}[H]
    \centering
        \includegraphics[width=3.0in]{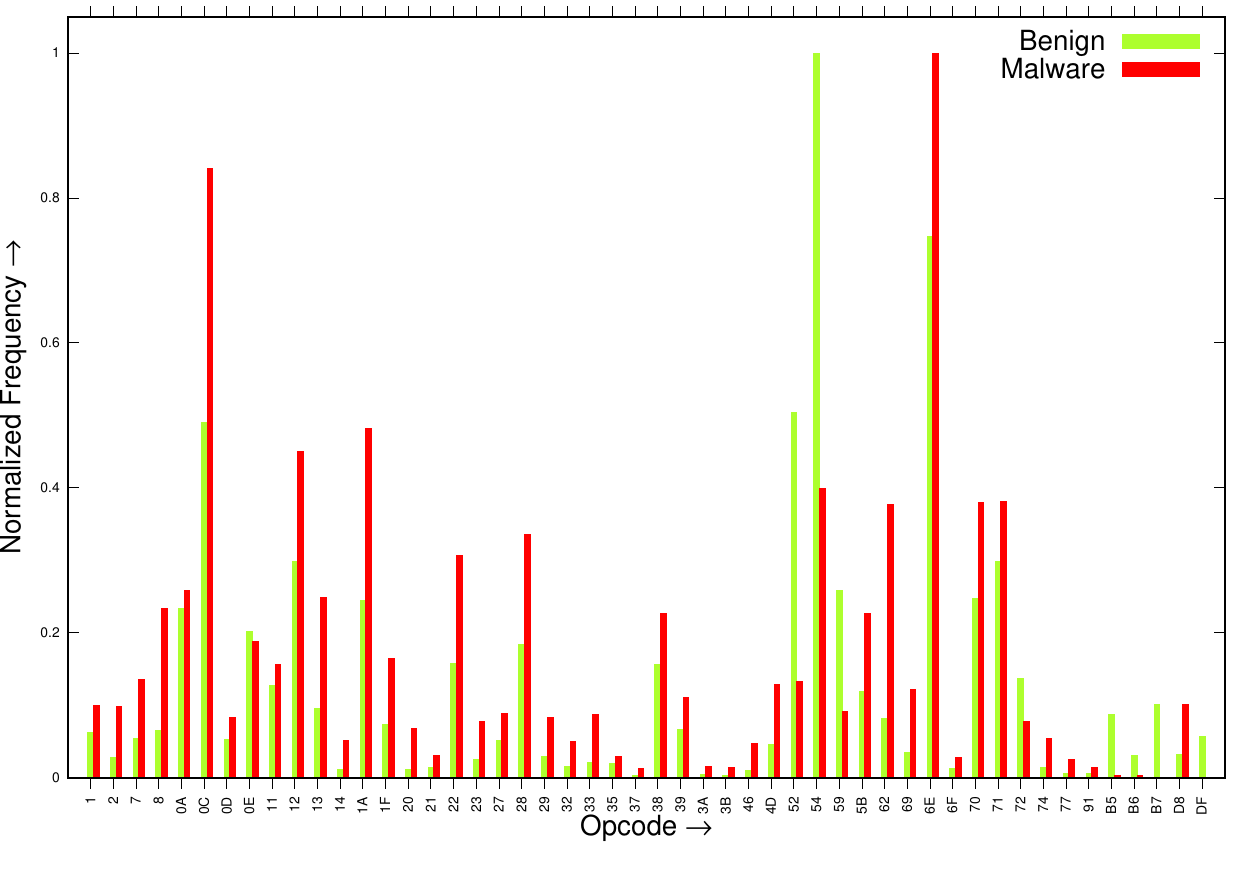} 
\caption{\small \sl Top 50 opcodes occurrence difference between benign and malicious apps in the OTHER group.} 
\label{fig:profileg6}
  \end{figure}

  We studied the occurrence of opcodes in benign and malicious apps separately in each formed group, 
and computed the  opcode occurrences difference between them. We observe that the opcode occurrence between malicious and benign apps among the formed group differ significantly (group-wise top 50 opcodes whose occurrence sign-

   \begin{figure}[H]
    \centering
        \includegraphics[width=3.0in]{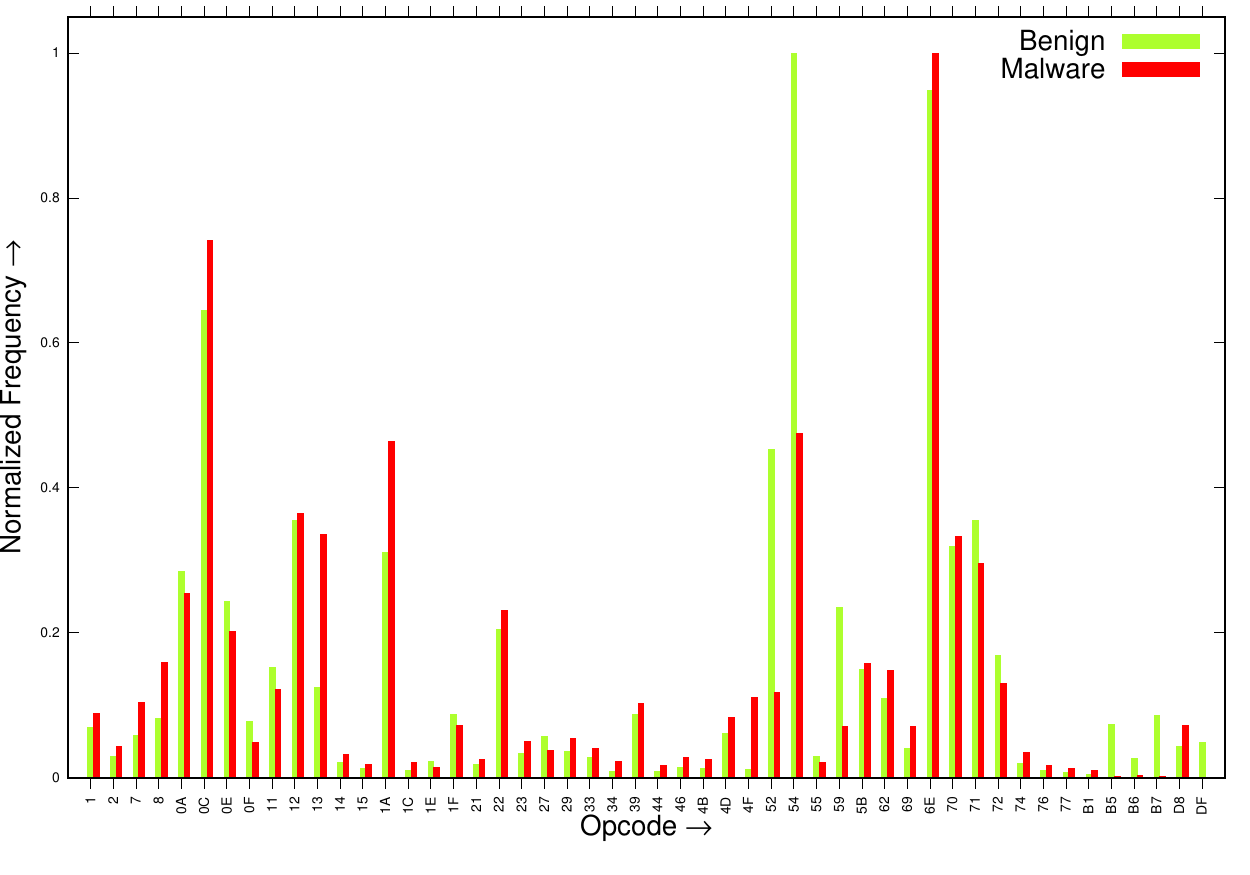} 
\caption{\small \sl Top 50 opcodes occurrence difference between benign and malicious apps in the Phone group.} 
\label{fig:profileg7}
  \end{figure}

   \begin{figure}[H]
    \centering
        \includegraphics[width=3.0in]{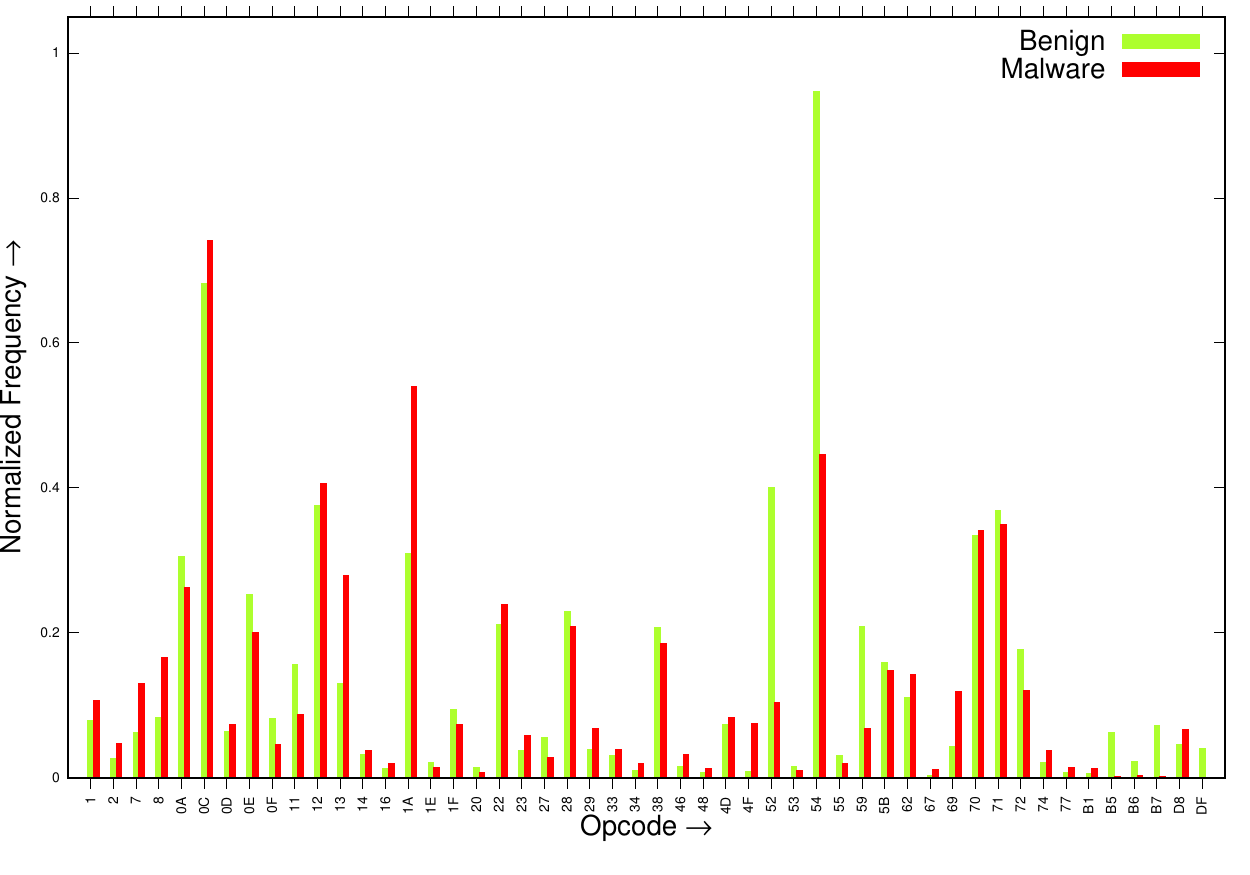} 
\caption{\small \sl Top 50 opcodes occurrence difference between benign and malicious apps in the SMS group.} 
\label{fig:profileg8}
  \end{figure}

   \begin{figure}[H]
    \centering
        \includegraphics[width=3.0in]{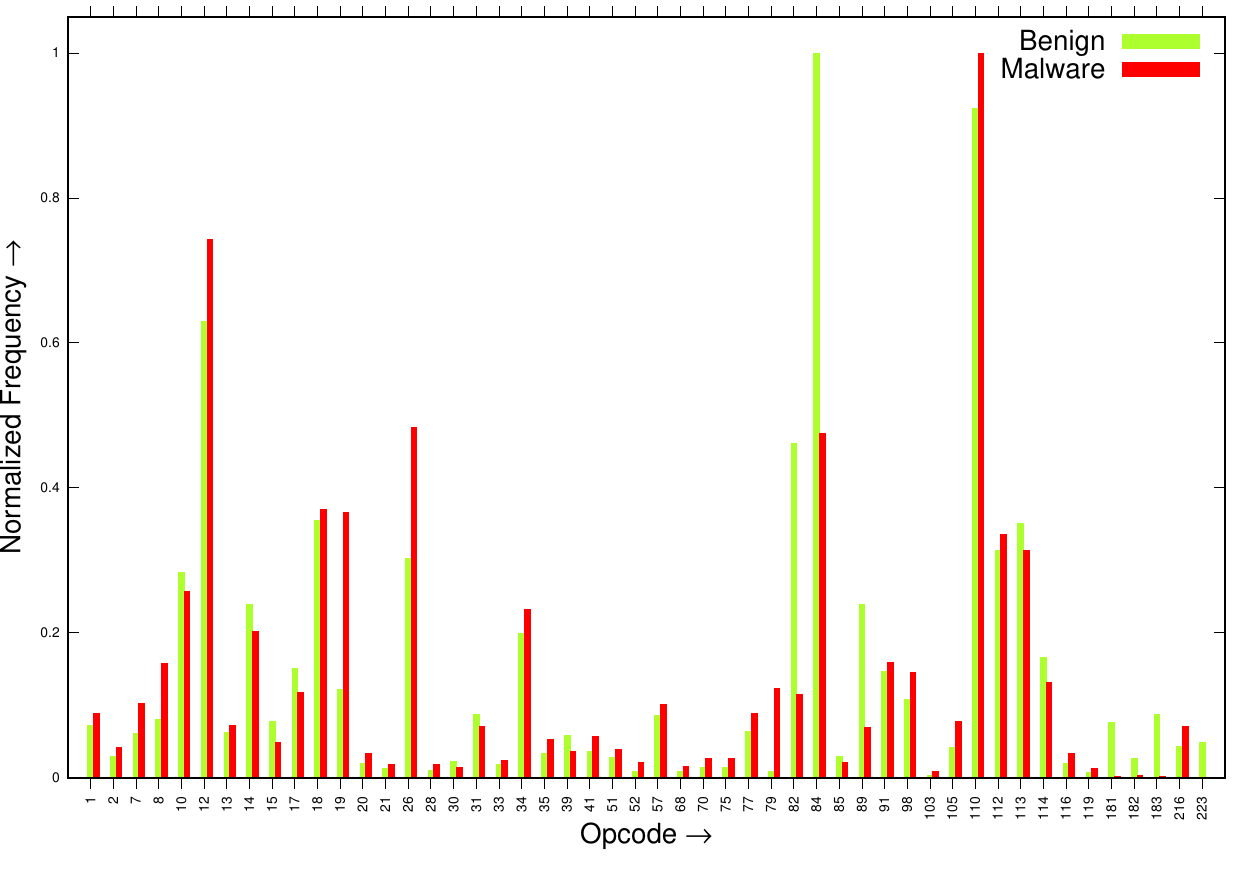} 
\caption{\small \sl Top 50 opcodes occurrence difference between benign and malicious apps in the Storage group.} 
\label{fig:profileg9}
  \end{figure}

 \noindent ificantly (group-wise top 50 opcodes whose occurrence significantly differ are shown in Figures \ref{fig:profileg1} - \ref{fig:profileg9} for the { \it Calendar, Camera, Contacts, Location, Microphone, Others, Phone, SMS,} and { \it Storage} group respectively). Also, we find that the opcode occurrence in any group differs significantly when compared with the opcode occurrence obtained without forming the groups \cite{sharmasahay}.
Hence, the final features are selected after ordering the opcodes by their occurrence difference in each group (Algorithm \ref{algo:FS}) and used it for the detection of Android malicious apps.

%
\begin{algorithm}[H]
\small

\medskip

\begin{flushleft}

\textbf{INPUT:} Pre-processed data\\
$\mathbf{N_B}$:  No. of benign apps, $\mathbf{N_M}$: No. of  malicious apps, 
\\$\mathbf{n}$: Total number of features required.
\\ \textbf{OUTPUT:} List of features
\end{flushleft}
\begin{algorithmic}

\STATE \textbf{BEGIN}

\FORALL{benign and malicious apps }
\STATE Find the sum of frequencies $\mathbf{f_i}$ of each opcode $\mathbf{Op}$  and normalize it. 
\STATE \begin{equation*} 
F_B ( Op_j ) = ( \sum f_i ( Op_j ) ) / N_B  
\end{equation*}
\STATE \begin{equation*} 
F_M ( Op_j ) = ( \sum f_i ( Op_j ) ) / N_M 
\end{equation*}
\ENDFOR

\FORALL{opcode $\mathbf{Op_j}$}
\STATE \begin{equation*} 
D(Op_j)= | F_B ( Op_j ) - F_M ( Op_j ) |
\end{equation*}
\ENDFOR

\RETURN $\mathbf{n}$ number of prominent opcodes as features with high $\mathbf{D(Op)}$.
\end{algorithmic}
\caption{\textbf{:} Feature Selection}
\label{algo:FS}
\end{algorithm}

   \begin{figure}[H]
    \centering
        \includegraphics[width=3.5in]{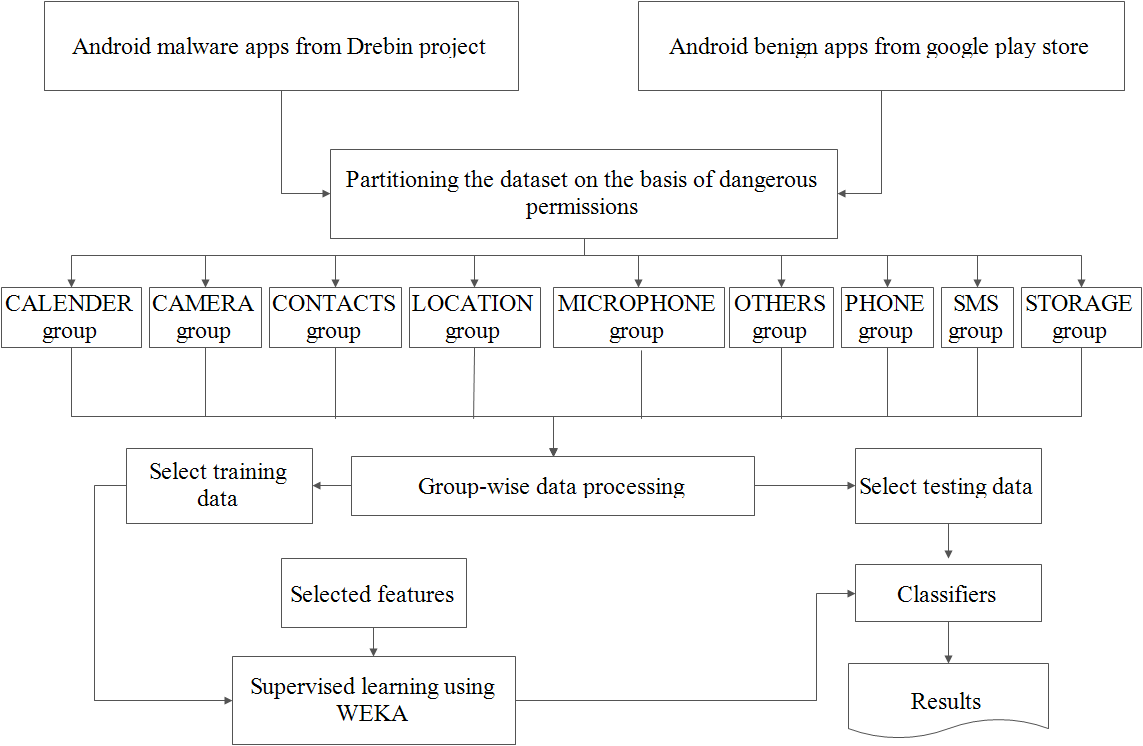} 
\caption{\small \sl Flow chart for the detection of Android malicious apps by grouping the data.} 
\label{fig:fc2}
  \end{figure}
  
\begin{table}[]
\centering
\small
\begin{tabular}{|l|c|c|c|c|c|}
\hline 
Groups & Train & Train & Test & Test & Total No.\tabularnewline
&malware&benign&malware&benign& of apps \tabularnewline
\hline 
Calendar & 59 & 57 & 14 & 14 & 144\tabularnewline
\hline 
Camera & 179 & 423 & 44 & 106 & 752\tabularnewline
\hline 
Contacts & 1073 & 356 & 268 & 89 & 1786\tabularnewline
\hline 
Location & 1538 & 68 & 383 & 18 & 2007\tabularnewline
\hline 
Microphone & 95 & 218 & 23 & 55 & 391\tabularnewline
\hline 
Others & 110 & 891 & 27 & 223 & 1251\tabularnewline
\hline 
Phone & 3981 & 1453 & 986 & 373 & 6793\tabularnewline
\hline 
SMS & 2712 & 239 & 677 & 60 & 3688\tabularnewline
\hline 
Storage & 2923 & 837 & 730 & 210 & 4700\tabularnewline
\hline 
\end{tabular}
\caption{Number of benign and Android malicious apps used for training and testing the classifiers.}
\label{table:nubfiles}
\end{table}

\section{Classification of Malicious Apps}
Ashu et al. \cite{sharmasahay} without grouping the data nor talking the apps permission investigated the top five classifiers viz. FT, RF, LMT, NBT and J48 for the classification of apps and reported that the FT is the best classifier and can detect the malicious apps with 79.27\% accuracy \cite{sharmasahay}. Hence to improve the detection accuracy in this paper, first we grouped the apps based on the permissions and then classify the malicious apps using prominent opcode as the features (Figure \ref{fig:fc2}). For the classification, the detail distribution (No. of training and testing malicious/benign apps, No. of apps in the group used for the classification) of the total collected dataset is given in Table \ref{table:nubfiles}. For the group-wise classification, we have used Waikato Environment for Knowledge Analysis (WEKA). 
On the  basis of studies \cite{sahay2016grouping} \cite{sharma2016improving}, we selected the same classifier (FT, RF, LMT, NBT, and J48) for the classification, but prominent features, training, and testing data are taken from the formed group only (Table \ref{table:nubfiles}). To measure the goodness of trained models, we evaluate the detection accuracy given by the equation
{\small
\begin{equation*}
\text{Accuracy}(\%) = \frac{\text{True Positive + True Negative}} {\text{Total No. of Android Apps}} \times 100
\end{equation*}
}
Where True Positive/Negative is the Android malicious/benign apps correctly classified \cite{sharmasahay}.	
\par
\begin{figure}[H]
    \centering
        \includegraphics[width=3.5in]{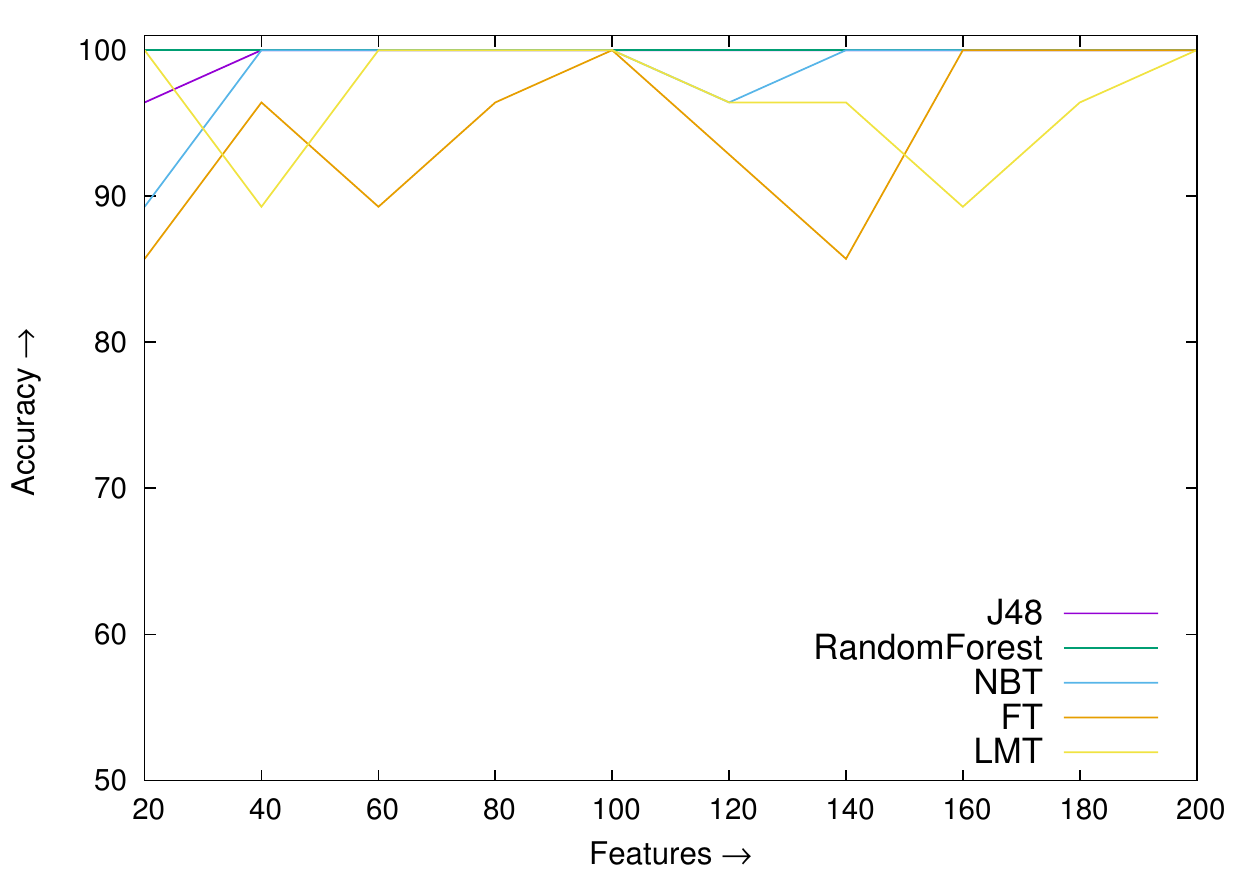} 
\caption{\small \sl Detection accuracy obtained by the selected five classifiers for the CALENDAR group.} 
\label{fig:f1}
  \end{figure}

\begin{figure}[H]
    \centering
        \includegraphics[width=3.5in]{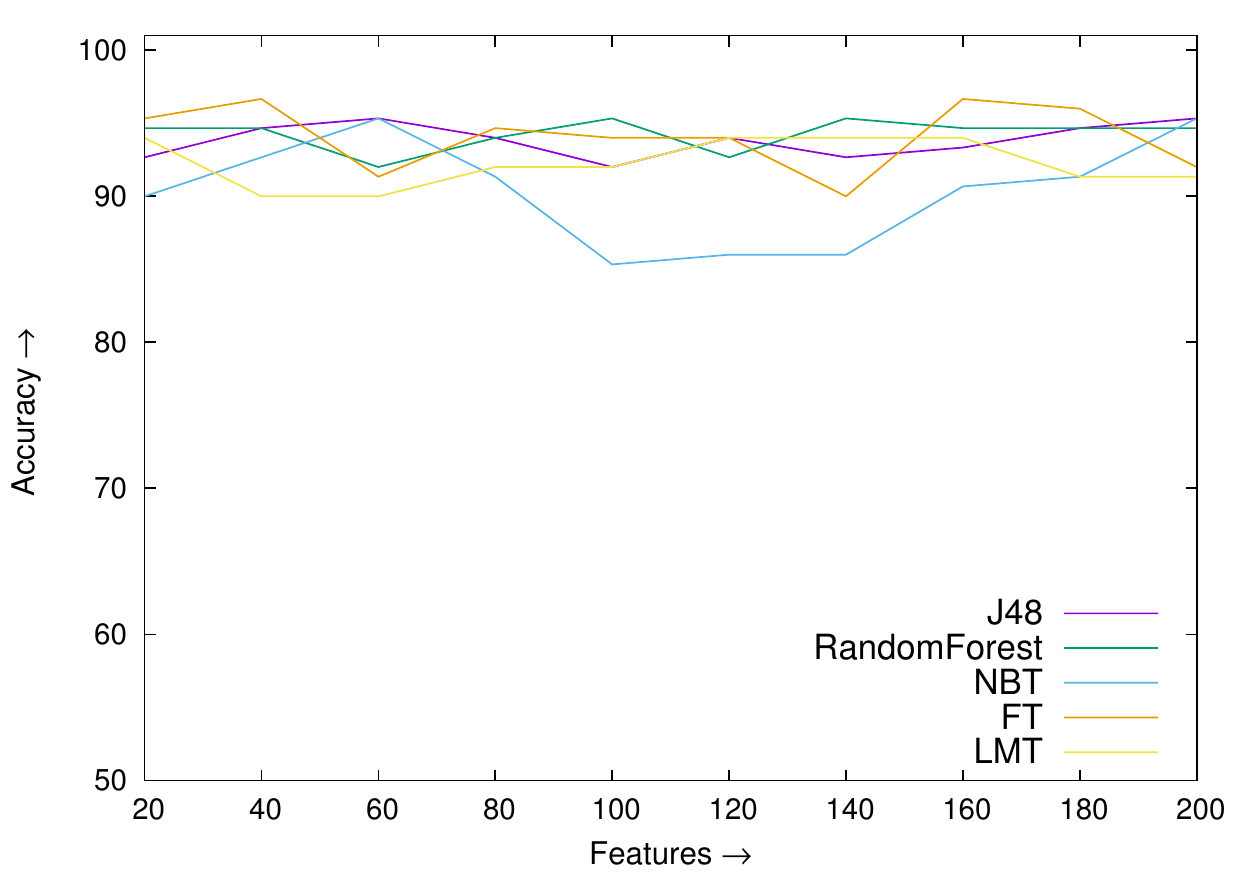} 
\caption{\small \sl Detection accuracy obtained by the selected five classifiers for the Camera group.} 
\label{fig:f2}
  \end{figure}

\begin{figure}[H]
    \centering
        \includegraphics[width=3.5in]{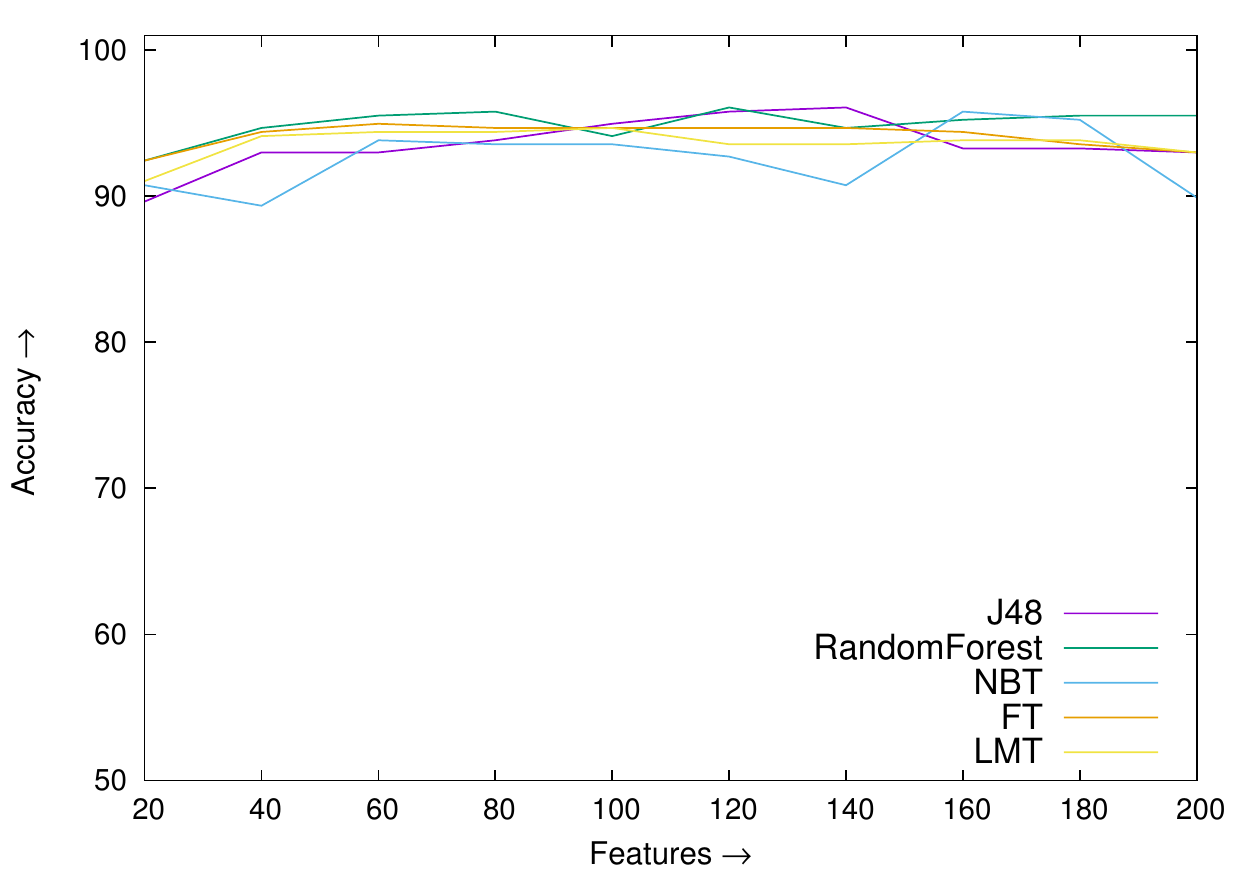} 
\caption{\small \sl Detection accuracy obtained by the selected five classifiers for the Contacts group.} 
\label{fig:f3}
  \end{figure}

\begin{figure}[H]
    \centering
        \includegraphics[width=3.5in]{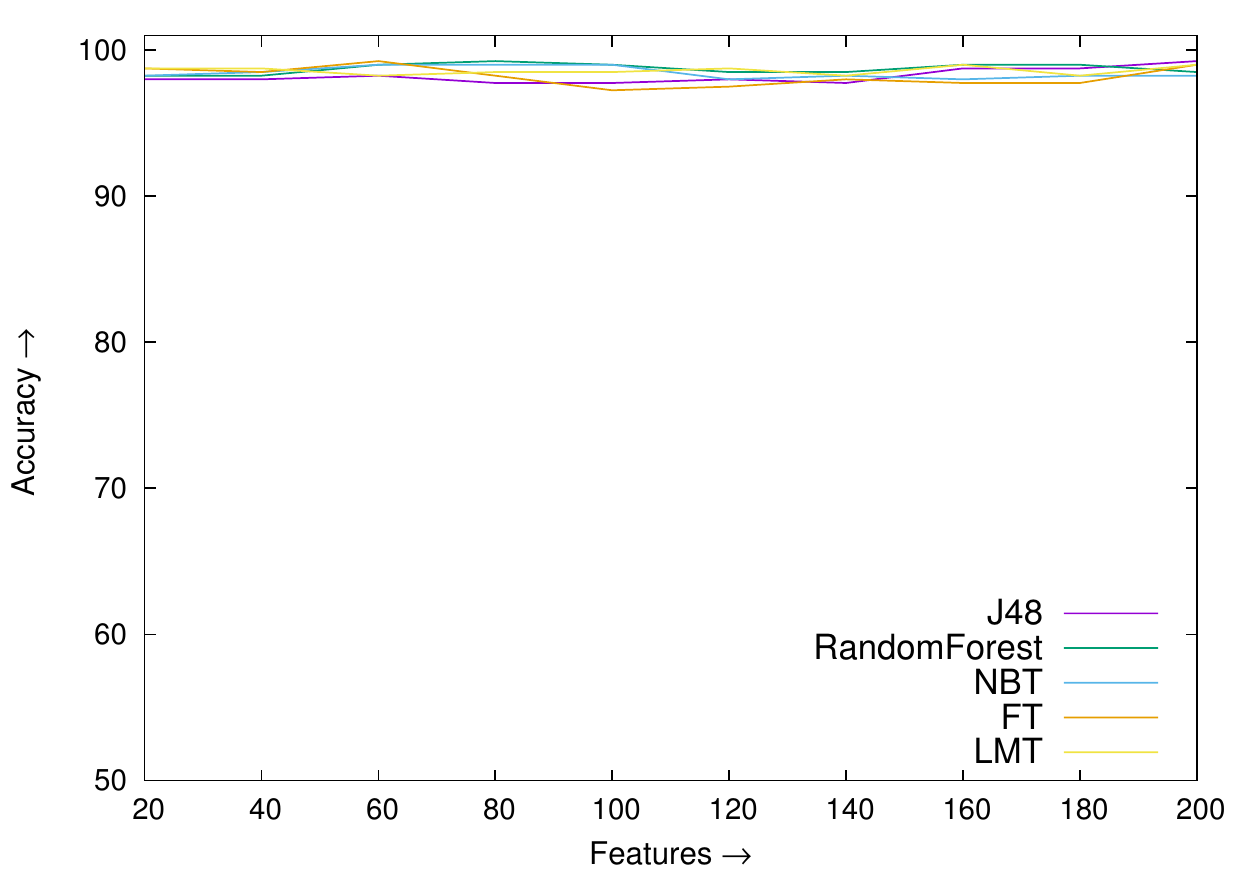} 
\caption{\small \sl Detection accuracy obtained by the selected five classifiers for the Location group.} 
\label{fig:f4}
  \end{figure}

\begin{figure}[H]
    \centering
        \includegraphics[width=3.5in]{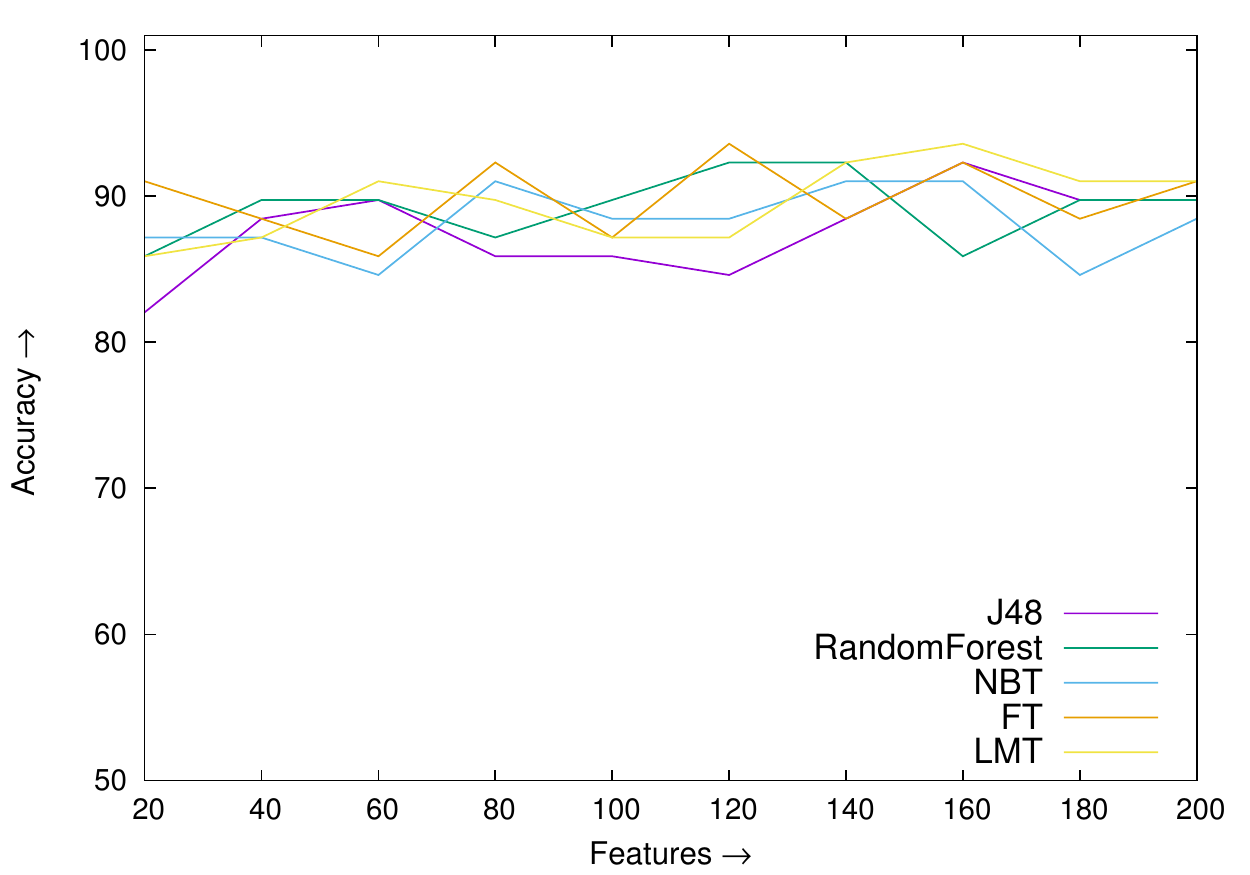} 
\caption{\small \sl Detection accuracy obtained by the selected five classifiers for the Microphone group.} 
\label{fig:f5}
  \end{figure}

\begin{figure}[H]
    \centering
        \includegraphics[width=3.5in]{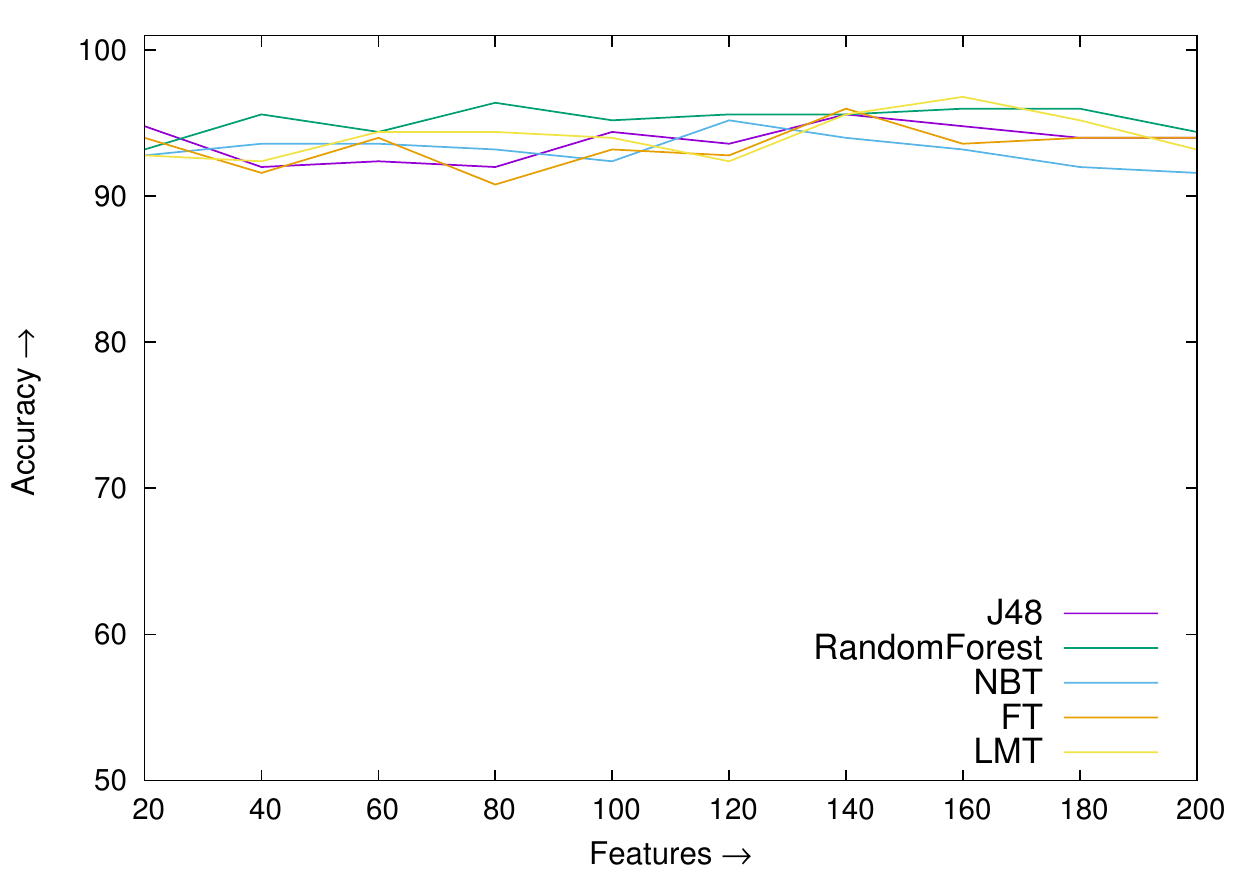} 
\caption{\small \sl Detection accuracy obtained by the selected five classifiers for the Others group.} 
\label{fig:f6}
  \end{figure}

\begin{figure}[H]
    \centering
        \includegraphics[width=3.5in]{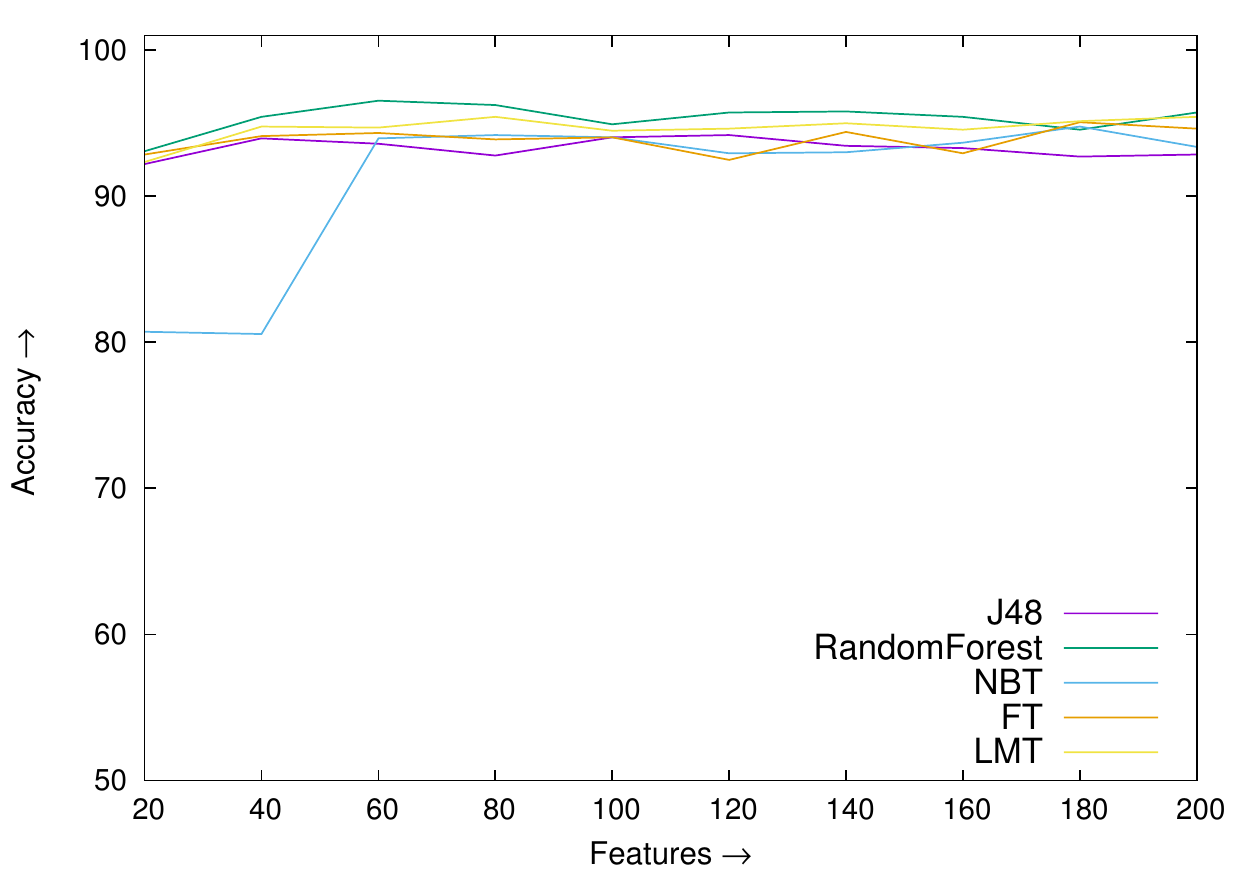} 
\caption{\small \sl Detection accuracy obtained by the selected five classifiers for the Phone group.} 
\label{fig:f7}
  \end{figure}

\begin{figure}[H]
    \centering
        \includegraphics[height=2.30in, width=3.5in]{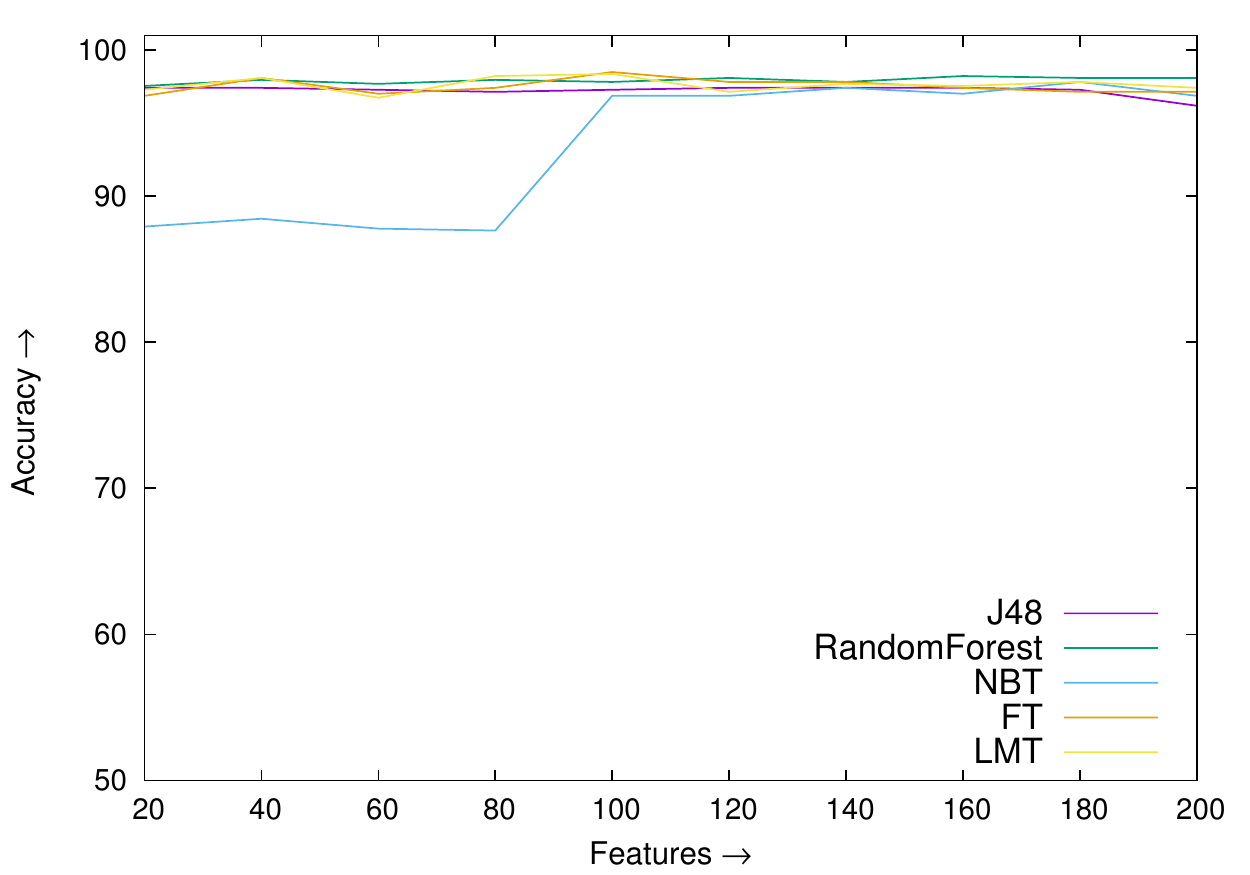} 
\caption{\small \sl Detection accuracy obtained by the selected five classifiers for the SMS group.} 
\label{fig:f9}
  \end{figure}

\begin{figure}[H]
    \centering
        \includegraphics[height=2.30in, width=3.5in]{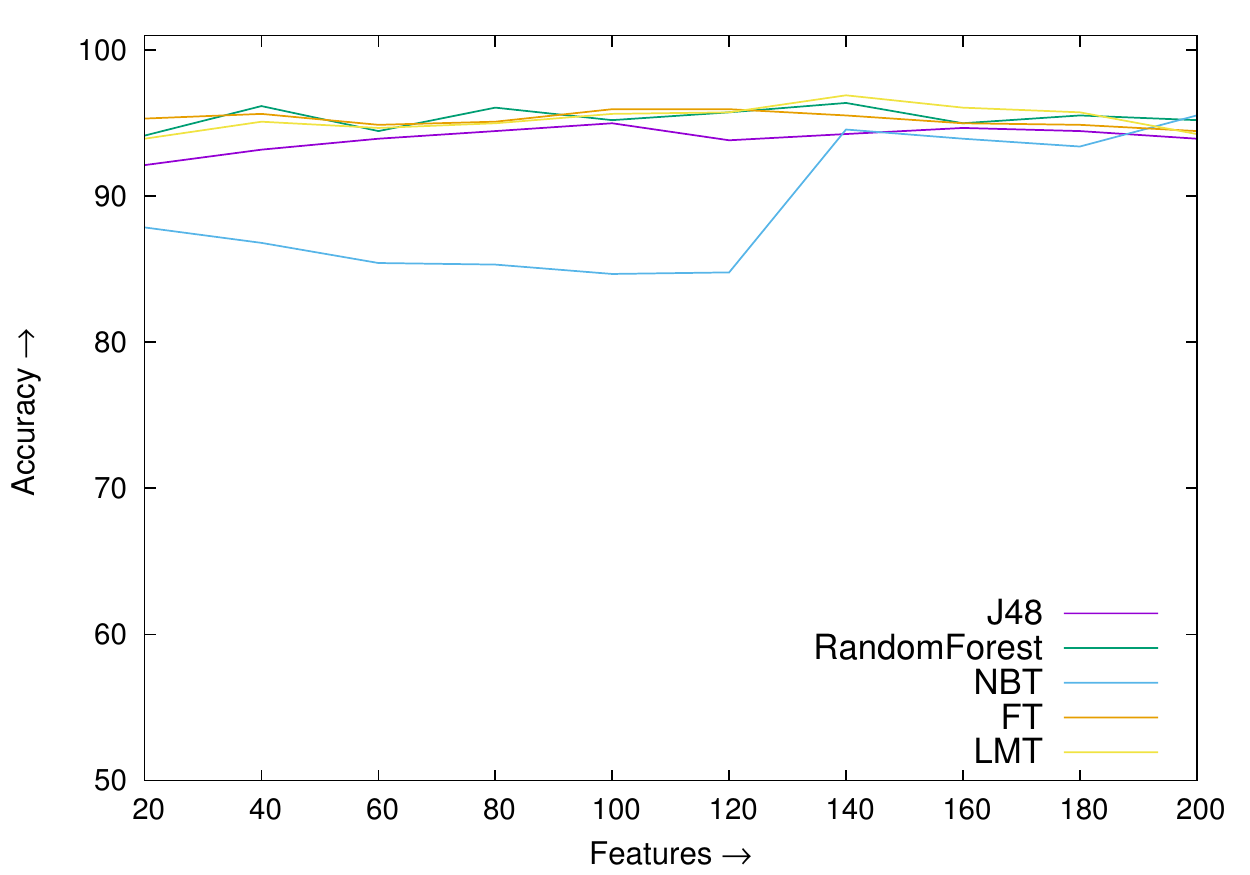} 
\caption{\small \sl Detection accuracy obtained by the selected five classifiers for the Storage group.} 
\label{fig:f10}
  \end{figure}

 The performance of the classifier has been investigated for each group by taking randomly 20\% of the collected data (other than the training) with 20 - 200 best features incrementing 20 features at each step and the result obtained are shown in Figures \ref{fig:f1} - \ref{fig:f10} for the {\it Calendar, Camera, Contacts, Location, Microphone, Others, Phone, SMS,} and {\it Storage} group respectively.
  
\begin{table}[H]
\centering
\small
\begin{tabular}{|c|c|c|c|c|c|} 
\hline 
No. of & J48 & RF & NBT & FT & LMT\tabularnewline
 Features & & & & & \tabularnewline
\hline 
20 & \color{light-gray}{\bf 93.69} & \color{light-gray}{\bf 95.01} & \color{light-gray}{\bf 90.37} & \color{light-gray}{\bf 93.32} & 94.28\tabularnewline
\hline 
40 & 95.28 & 96.26 & 92.26 & 93.78 & \color{light-gray}{\bf 93.45}\tabularnewline
\hline 
60 & 95.51 & 96.10 & 94.24 & 94.01 & 94.31\tabularnewline
\hline 
80 & 94.83 & \color{dark-gray}{\bf 96.32} & 94.44 & 95.38 & 95.46\tabularnewline
\hline 
100 & \color{dark-gray}{\bf 95.15} & 96.24 & 94.41 & \color{dark-gray}{\bf 95.43} & \color{dark-gray}{\bf 85.47}\tabularnewline
\hline 
120 & 94.48 & 95.96 & 92.96 & 94.57 & 94.23\tabularnewline
\hline 
140 & 95.12 & 96.08 & 93.68 & 93.53 & 94.76\tabularnewline
\hline 
160 & 95.39 & 95.16 & \color{dark-gray}{\bf 94.97} & 95.16 & 94.29\tabularnewline
\hline 
180 & 94.94 & 95.73 & 93.93 & 95.18 & 94.56\tabularnewline
\hline 
200 & 94.71 & 95.78 & 93.24 & 94.98 & 94.71\tabularnewline
\hline 
\color{dark-gray}{\bf Maximum }& \color{dark-gray}{\bf 95.51} & \color{dark-gray}{\bf 96.32} & \color{dark-gray}{\bf 94.97} & \color{dark-gray}{\bf 95.43} & \color{dark-gray}{\bf 95.47}\tabularnewline
\hline 
 \color{light-gray}{\bf Minimum} &  \color{light-gray}{\bf 93.69} & \color{light-gray}{\bf  95.01} &  \color{light-gray}{\bf 90.37 }&  \color{light-gray}{\bf 93.32} &  \color{light-gray}{\bf 93.45}\tabularnewline
\hline 
\end{tabular}
\caption{Average accuracy obtained by the five classifiers.}
\label{table:avg_accuracy}
\end{table}

\begin{figure}[H]
    \centering
        \includegraphics[height=1.970in, width=3.2in]{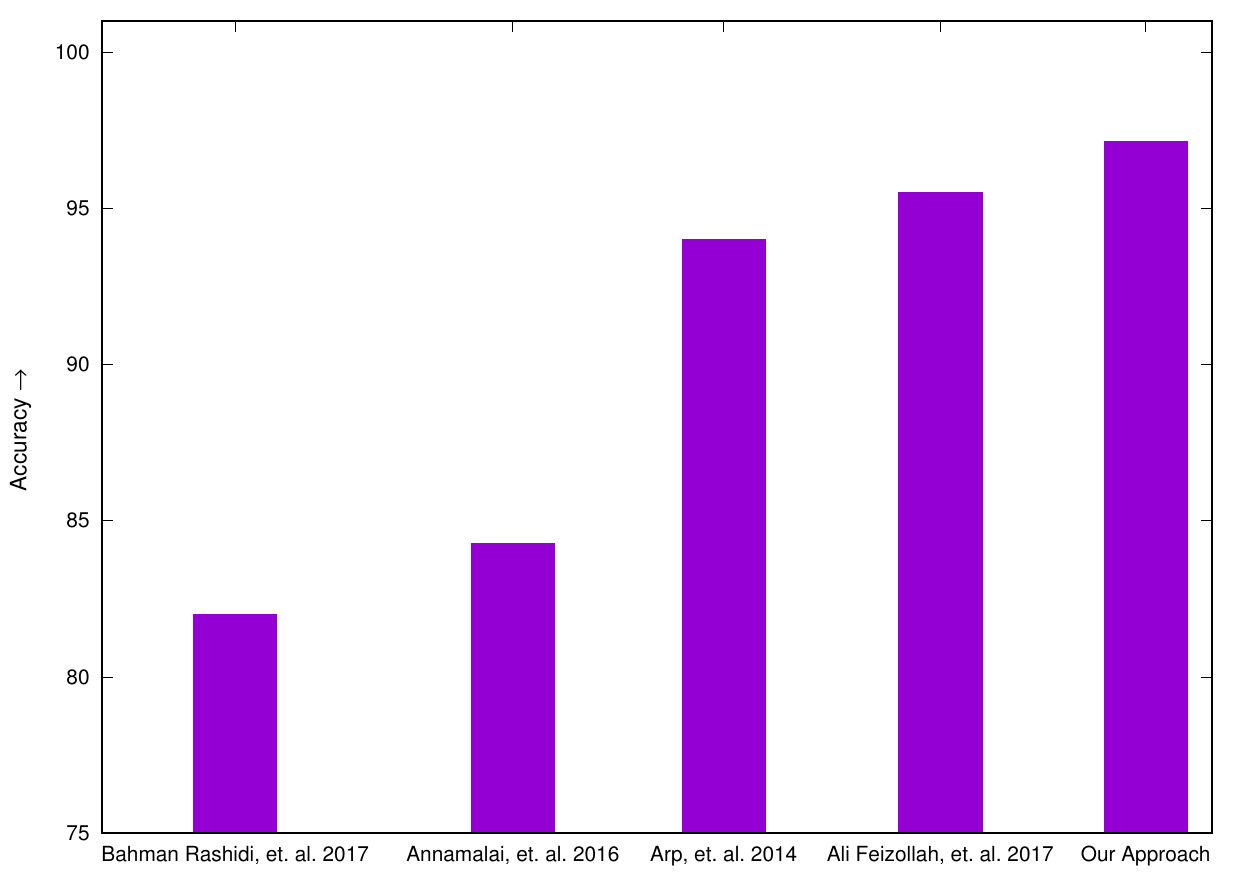} 
\caption{\small \sl Comparisons of accuracy achieved by us with four other authors.} 
\label{fig:comp}
  \end{figure}

The average accuracy obtained by the selected classifier are shown in Table \ref{table:avg_accuracy}. Here, the average accuracy means the sum of accuracy obtained by the classifier in the individual group with a fixed number of features divided by the total number of groups. 
The analysis shows that RF average detection accuracy is best among the five classifiers and fluctuates least with the number of features, whereas NBT performance is worst and fluctuate maximum with the number of features.
However, the maximum average accuracy obtained by the selected five classifiers does not fluctuate much (94.97\% - 96.32\%) but minimum average accuracy fluctuation is high (90.37\% - 95.01\%), and for the best performance one shall take top 80 - 100 features, for the training and testing.
The best accuracy obtained by the classifier in all the groups are given in Table \ref{table:max_accuracy}.
We find that the detection accuracy is maximum in the Calendar group and minimum in the Microphone group obtained by FT and RF classifier respectively.
The overall average maximum accuracy comes to 97.15\%, which is very much better than then the obtained accuracy without grouping and taking permissions into account \cite{sharmasahay} and Arp, et. al. (94\%, 2014), Annamalai et. al. (84.29\%, 2016), Bahman Rashidi et. al. (82\%, 2017), Ali Feizollah, et. al. (95.5\%, 2017) (Figure \ref{fig:comp}). 
In terms of $TP$ i.e. detection rate of malicious apps, the {\it Calendar} group are best classified by RF and {\it SMS} group are least by FT, while in terms of $TN$ i.e. benign detection rate, {\it Calendar}, and {\it SMS} group are best classified with RF and FT classifier respectively, while Others group containing normal permissions is best classified by the LMT classifier. The group-wise results of TP and TN obtained by the classifiers which give the best accuracy are shown in Table \ref{table:max_accuracy}.

\begin{table}[]
\centering
\small
\begin{tabular}{|c|c|c|c|c|c|}
\hline \centering
Groups &  Best & Accu- &  Features & TN & TP\tabularnewline
 &  Classifier &racy & Required &  & \tabularnewline
\hline 
Calendar & RF & 100.00 & 20 & 1.00 & 1.00\tabularnewline
\hline 
Camera &  FT & 96.67 & 40 & 0.93 & 0.98\tabularnewline
\hline 
Contacts & RF & 96.08 & 120 & 0.99 & 0.89\tabularnewline
\hline 
Location &  FT & 99.25 & 60 & 0.99 & 0.94\tabularnewline
\hline 
Microphone &  FT & 93.59 & 120 & 0.87 & 0.96\tabularnewline
\hline 
Others &  LMT & 96.80 & 160 & 0.85 & 0.98\tabularnewline
\hline 
Phone & RF & 96.54 & 60 & 0.98 & 0.92\tabularnewline
\hline 
SMS &  FT & 98.51 & 100 & 1.00 & 0.80\tabularnewline
\hline 
Storage &  LMT & 96.91 & 140 & 0.99 & 0.88\tabularnewline
\hline 
\end{tabular}
\caption{Group-wise maximum accuracy, TP and TN obtained by the classifiers.}
\label{table:max_accuracy}
\end{table}

 \section{Conclusion}
For the smart devices users, millions of Android apps are available at Google Play store and by  the third party. Some of these available apps may be malicious. To defend the threat/attack from these malicious apps, a timely counter-measures has to be developed.
Therefore, in this paper using \textit{Drebin} benchmark malware dataset we group-wise analyzed the collected data based on permissions and experimentally demonstrated how to improve the detection accuracy of Android malicious apps and achieved 97.15\% average accuracy.  The obtained results outperformed
 the accuracy achieved by without grouping the data (79.27\%, 2016), Arp, et al. (94\%, 2014), Annamalai et al. (84.29\%, 2016), Bahman Rashidi et al. (82\%, 2017)) and Ali Feizollah, et al. (95.5\%, 2017). The outperformance of our approach  with the compared author results is basically due to the use of logic of the apps resides in the {\it .smalli} file and developing nine different
 models for the classification. Among the groups, the {\it Microphone} group detection accuracy is least while {\it Calendar} group apps are detected with maximum accuracy and for the best performance, one shall take top 80 - 100 features. In term of TP i.e. detection rate of malicious apps, {\it Calendar} group is best classified by RF, and {\it SMS} group is least by FT, while in terms TN i.e. benign detection rate, {\it Calendar}, and {\it SMS} group are best classified by RF and FT classifier respectively, while Others group containing normal permissions is best classified by the LMT classifier. It appears that group-wise detection of Android malicious apps will be efficient than without grouping the data. Hence, for the efficient classification of apps, in-depth study is required to optimize the feature selection, identifying the best-suited classifier for the group-by-group analysis. In this direction, work is in progress and will be reported elsewhere.

%
\section*{Acknowledgements} 
Mr. Ashu Sharma is thankful to BITS, Pilani, K.K. Birla Goa Campus for the Ph.D. scholarship No. Ph603226/July 2012/01. We are also thankful to Technische Universitat Braunschweig for providing the Drebin dataset for research on Android malware.

\addcontentsline{toc}{chapter}{\protect\numberline{}{REFERENCES}}
\bibliography{ijns}
\bibliographystyle{plain}

\section*{Biography}
\noindent {\bf Mr. Ashu Sharma} was born in Jhansi, Uttar Pradesh, India. He received his  Bachelor's degree in Computer Science and Engineering from Uttar Pradesh Technical University and Master's degree in Information Security from Atal Bihari Vajpayee Indian Institute of Information Technology and Management, Gwalior. In 2012 he joined the Department of Computer Science and Information Systems, BITS, Pilani, K.K. Birla Goa Campus, India as a full-time research scholar for the Ph.D. degree under the supervision of Dr. Sanjay K. Sahay. He has published several papers in reputed journals and conferences.
\vspace*{0.2cm}\\
\noindent {\bf Dr. Sanjay Kumar Sahay} is working as an Associate Professor in the Department of Computer
Science and Information Systems, BITS, Pilani, K.K. Birla Goa Campus. He is also a Visiting Associate of IUCAA, Pune. His research interests are Information Security, Data Science, and Gravitational Waves. He basically teaches Network Security, Cryptography, Computer Networks, and Data Mining courses. Before joining BITS, Pilani, and after submitting his Ph.D. thesis on ``Studies in Gravitational Wave Data Analysis" during 2002-2003, he continued his work on Data Analysis of Gravitational Waves as a Project Scientist at IUCAA, Pune, India. In 2003-2005 at Raman Research Institute, Bangalore, India he worked as Project Associate on the multi-wavelength astronomy project (ASTROSAT), where he worked on the data pipeline of Scanning Sky Monitor. In 2005 he worked as Post Doctoral Fellow at Tel Aviv University.  \vspace*{0.2cm}\\

\end{document}